\documentclass[conference]{IEEEtran}

\usepackage{setspace}
\usepackage{url}
\usepackage{amssymb, amsmath}
\usepackage{graphicx,subfigure}
\usepackage{times}
\usepackage{color}
\usepackage{multirow}
\usepackage{epstopdf}

\newcommand{\BB}[1]{\Bigl[ #1\Bigr]} 
\newcommand{\bp}[1]{\bigl( #1\bigr)} 
\begin{document}
\title{Battery-Aware Relay Selection for Energy Harvesting Cooperative Networks}
\author{
\IEEEauthorblockN{Yu-Hsien Lee and Kuang-Hao Liu}\\
\IEEEauthorblockA{Institute of Computer and Communication Engineering\\
National Cheng Kung University\\
Tainan, Taiwan 701\\
Email: \{q36014413, khliu\}@mail.ncku.edu.tw} }
\date{}

\graphicspath{{./figure/}}

\maketitle

\begin{abstract}
The use of energy harvesting (EH) nodes as cooperative relays is an emerging
solution for enabling green wireless systems. In this paper, we consider
multiple EH relay nodes harvesting energy from the radio frequency (RF) signal
received from the source and use that harvested energy to forward the source
information to the destination. Unlike conventional relays with fixed power supplies, EH relays may not be permanently available to assist the
source transmission due to the limited energy conversion efficiency, the
mismatch between the charging and discharging profiles, and the finite energy
storage capacity. We propose the battery-aware relay selection (BARS) scheme,
which jointly considers the channel condition and the battery status for relay
selection. The outage probability of the proposed scheme is analyzed using a
Markov chain model. Simulations are performed to validate the analysis
accuracy. Through numerical results, we show that the proposed BARS scheme can
achieve full diversity order equal to the number of relays without the need of
fixed power cables.
\end{abstract}

\section{Introduction}
Powering wireless devices by sustainable energy is an emerging solution to
enable green wireless networks~\cite{Cai2011}. Known as energy harvesting (EH),
passively powered devices collect energy from external power sources, such as
vibration, solar, thermoelectric effects, ambient radio frequency (RF)
radiation, and so forth, to maintain their physical operations without any
wiring cost. In this work, we are interested in EH based on RF
signals~\cite{Ju2014}.

One of the potential applications of EH nodes is the cooperative relays, which
are deployed to extend network coverage and improve transmission reliability
between two distant nodes. Traditionally, relays are powered by fixed power
supplies, leading to extra power consumption for information relaying. It is
thus desirable to replace traditional relays by EH relays that power themselves by the
energy harvested from the source signal as a green communication solution. In
this context, no additional power is consumed to perform information relaying
but the key challenge is that these EH relays may not be permanently available
to help as their traditional counterparts. When more EH relays are short of
enough energy to transmit, it implies less diversity branches
can be used to pass the source information, leading to low diversity gain.

Several practical constraints hinder the EH relays from being useful to
cooperate. Firsly, only a portion of the harvested energy is available to use,
because the energy collected by the energy harvester circuit needs to be
converted to DC voltage first. Depending on the conversion circuit design, the
energy conversion efficiency reported in the literature varies from $30\%\sim
50\%$~\cite{Le2008a,Penella-Lopez2011}. Since the harvested energy from a
single shot may be far from enough to be used for transmission, it is desirable
to accumulate the harvested energy by storing it in an energy storage such as a
rechargable battery or a super cpacitor for the later use. In practice, the
energy storage is limited in size, and thus EH relays may encounter energy
shortage whenever the energy consumption rate is higher than the energy
harvesting rate. One countermeasure is thus to select those relays with
sufficient energy to cooperate via a certain relay selection scheme.

Relay selection has been extensively addressed for conventional relays.
Previous research indicates that selecting one relay with the superior channel
condition than the others is promising in achieving the same
diversity-multiplexing tradeoff as that by using sophiticated space-time coding
schemes~\cite{Bletsas2006,Beres2006}. Such a relay selection scheme, referred
to channel state information (CSI)-based scheme, may fail to fully exploit the
diversity gain if the selected relay lacks of sufficient power to transmit and
thus it is not suitable to cooperative networks with EH relays.
In~\cite{Liu2014}, the CSI-based relay selection is applied to cooperative
networks where EH relays are subject to finite energy storage and limited
energy conversion efficiency. It is shown that the CSI-based relay selection
scheme does not achieve any diversity gain even with a large battery.
In~\cite{Ding14}, the authors consider two relay selection schemes, namely the
random relay selection and the distance-based relay selection schemes, in
cooperative networks with EH relays. Their analysis shows that the diversity
gain achieved by these two methods is at most two. The analysis conducted
in~\cite{Ding2014b} further reveals that the diversity gain achieved by the
CSI-based relay selection scheme using EH relays is only half of that using
traditional relays.

In this work, we propose a new relay selection for EH relays. The proposed
scheme, referred to as battery-aware relay selection (BARS), employs both CSI
and battery status for making the relay selection decision. We analyze the
outage probability of BARS by developing a Markov-chain model that captures the
evolution of battery status at each relay selection epoch. Numerical results
are presented to validate the analysis accuracy and demonstrate the performance
of the proposed relay selection scheme with EH relays subject to numerous
system parameters. The rest of this paper is organized as follows.
Sec.~\ref{sec:model} explains the system model. The traditional CSI-based and
the proposed relay selections schemes are introduced in Sec.~\ref{eq:proposed}.
Performance analysis is conducted in Sec.~\ref{sec:analysis}, followed by
numerical results in Sec.~\ref{sec:results}. Concluding remarks are drawn in
Sec.~\ref{sec:conclusion}.

\begin{figure}[!t]
\centering
\includegraphics[width=0.7\columnwidth]{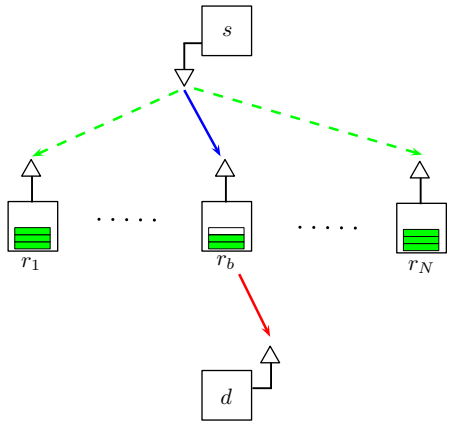}
\caption{A two-hop cooperative communication network with multiple relays
powered by the energy harvested from the source signal.} \label{fig:model}
\end{figure}
\section{System Model}\label{sec:model}%
Consider a multi-relay network with one source $s$, one destination $d$, and
$N$ relays $r_1,\cdots,r_N$, as shown in Fig.~\ref{fig:model}. The
communication between $s$ and $d$ relies on the intermediate relays that
perform decode-and-forward (DF) to forward the source information, assuming that the
direct link between $s$ and $d$ is not available. In this work, the RF signal emitted by $s$ is the sole energy source for relays. The harvested energy is stored in a rechargeable battery of size
$B$ by converting the RF signal into the DC voltage. The conversion efficiency
is characterized by the parameter $\kappa \in [0,1]$. The battery size is
assumed to be identical for all relays, and a discrete battery model is
employed. Specifically, each relay battery is quantized into $L$ levels. Let
$b_l$ denote the $i$th quantization level for $l=0, 1, \cdots, L+1$. The battery
level associated with relay $r_i$, denoted as $V_i$, corresponds to level $l$
if $V_i \in (b_l, b_{l+1}]$ with $b_0 = 0$ and $b_{L+1} = B$. The battery is
assumed to be linear such that the charging and discharging rates are constant.

The source transmit with a fixed power $P$, while relay $r_i$ transmit power
$P_{r_i}$ is adjusted to ensure the successful decoding at the destination. To support the transmission rate of $R$ bits/sec/Hz, the decoding threshold $T$ equals $2^{2R}-1$. Hence the minimum transmit power required for
successful decoding by $d$ is $P_{r_i}=T/h_{i}$, where $h_i$
denotes the channel gain power between $r_i$ and $d$. The harvested energy from
the source transmission at relay $r_i$ is $E_i=P g_i \kappa$, where $g_i$ is
the channel gain power between $s$ and $r_i$. Assuming Rayleigh fading
channels, both $g_i$'s and $h_i$'s are exponentially distributed with mean
$\bar{g}_i$ and $\bar{h}_i$, respectively. In addition, the thermal noise power
$N_0$ at the receiving end is assumed to be identical for all nodes.
Therefore, the signal-to-noise ratio (SNR) $P/N_0$ is a constant.

\section{Relay Selection}\label{eq:proposed}
We first review the CSI-based relay selection scheme for
conventional relays. Considering DF relays, define the set of relays
that can decode successfully as the decoding set, i.e.,
$\mathcal{D}(s)=\{ r_i | \tfrac{P_s g_i}{N_0} \geq T\}$. From this
decoding set, the best relay, denoted as $r_b^{\text{CSI}}$, is
chosen as the one with the superior relay-destination channel
condition than the others. Such a CSI-based relay selection scheme
can be expressed as
\begin{equation}\label{eq:csi}
r_b^{\text{CSI}} = \arg~\max_{ r_i \in \mathcal{D}(s) } h_i.
\end{equation}
This is scheme is recognized as the optimal diversity-achieving
relay selection scheme because it achieves the same diversity gain
as using multiple relays to forward the source signal but consuming
much less radio resources. For EH relays with finite energy storage,
however, the selected relay according to (\ref{eq:csi}) might lack
of enough power to transmit, yet its channel condition is the best.

To overcome this drawback, we propose to select the cooperating
relay with battery status taking into account. To this end, we first
define a subset of relays that not only can decode the source
information but also has sufficient power to transmit. Such a set of
relays is referred to as the forwarding set defined as
\begin{equation}\label{eq:F(s)}
\mathcal{F}(s) = \left\{ r_i \Big| ( \frac{P_sg_i
}{N_0} \geq T ) \cap (V_i \geq P_{r_i} ) \right\},
\end{equation}
where $V_i$ is the current battery level of relay $r_i$ at the selection epoch.
Given the forwarding set, the best relay selected by the proposed scheme,
referred to as battery-aware relay selection (BARS), can be expressed as
\begin{equation}\label{eq:bars}
r_b^{\text{BARS}} = \arg~\min_{ r_i \in \mathcal{F}(s) } E_i,
\end{equation}
where $E_i = P_s g_i \kappa$ is the harvested energy by relay $r_i$. The
rational behind BARS is two-fold. Firstly, selecting the best relay from the
forwarding set ensures that the selected relay can successfully decode the
source information and has sufficient power to transmit. This is important to
fully exploit the selection gain provided by multiple relays based on energy
harvesting. Secondly, by choosing the relay with the minimum harvested energy
in the forwarding set, the accumulated amount of harvested energy per source
transmission is maximized because there are always at least $N-1$ relays
performing energy harvesting (it is possible that all the $N$ relays will harvest energy if the
forwarding set is empty).

We note that although we do not have a rigorous proof for the achievable diversity gain of BARS, the numerical results shown in Sec.~\ref{sec:results} reveal that BARS is plausible to
achieve full diversity order equal to the number of available relays. The key
to this success lies in the consideration of the forwarding set $\mathcal{F}(s)$ in
relay selection. If we replace the role of $\mathcal{F}(s)$ in the selection
rule (\ref{eq:bars}) by $\mathcal{D}(s)$, i.e., ignoring the battery status,
full diversity gain is not guaranteed. The modified scheme,
referred to the benchmark scheme, can be expressed as
\begin{equation}\label{eq:benchmark}
r_b^{\text{Benchmark}} = \arg~\min_{ r_i \in \mathcal{D}(s) } E_i,
\end{equation}
and its performance will be discussed in Sec.~\ref{sec:results}.

\section{Performance Analysis}\label{sec:analysis}
In BARS, the outage event occurs only when all the relays are in the harvesting
mode, i.e, the forward set $\mathcal{F}(s)$ is empty. From (\ref{eq:F(s)}),
whether a relay performs energy harvesting or data forwarding depends on
both its battery status and the channel conditions.
Based on the discrete battery model, the battery of an arbitrary relay may be
in one of the $L+2$ levels and there are total of $(L+2)^N$ combinations of
battery status for $N$ relays. Denote $s_j = (V_1, V_2, \cdots, V_N)$ as the
$j$th combination, where $V_i=\{0,\cdots,L+1\}$ for $i=1,\cdots,N$. The outage probability of BARS
can be expressed in the following general form as
\begin{equation}\label{eq:pout}
P_{\text{out}} = \sum_{j=1}^{(L+2)^N} \Pr[\mathcal{F}(s) = \emptyset | s_j
]\Pr[s_j].
\end{equation}
In~(\ref{eq:pout}), the conditional probability, $\Pr[\mathcal{F}(s) = \emptyset | s_j
]$ depends on the specific configuration of $s_j$ and thus there is no general form. On the other hand,  the probability $\Pr[s_j]$ can be obtained by modeling the charging/discharging behavior of each relay battery status as a discrete-time Markov chain (DTMC) with finite states.
%
%
The transition probability of the DTMC is defined as $\mathbf{P}=[p_{j,k}]$
where $p_{j,k}$ denotes the transition probability from state $s_j$ to state
$s_k$. It can be verified that $\mathbf{P}$ is irreducible and row stochastic, and thus
there exists an unique steady-state probability vector
$\boldsymbol{\pi}=\{\pi_j \}_{j=1}^{(L+2)^N}$, where $\pi_j=\Pr[s_j]$. Again,
$\mathbf{P}$ does not have a general expression but can be obtained
explicitly given the numbers of relays and battery levels. In the following, we
first derive the battery state transitions of an arbitrary relay, which serve
as the basis for constructing $\mathbf{P}$. To ease the presentation,
$F_X(\cdot)$ represents the CDF of a random variable $X$.

\subsection{Battery State Transitions of An Arbitrary Relay}
For convenience, define $A_f(m)$ the event that a relay with battery state
$V_m$ is in the forwarding mode. According to (\ref{eq:F(s)}), $A_f(m)
\triangleq (P g_i/N_0 \geq T) \cap (b_m \geq P_{r_i})$ with probability
\begin{equation}\label{eq:P_Af}
\Pr[A_f(m)] = (1-F_{g_i}\bp{ \tfrac{TN_0}{P} } ) (1-F_{h_i}\bp{ \tfrac{T}{b_m}
} )
\end{equation}
Similarly, denote the event of a relay with battery state $V_m$ is in the
charging mode mode by $A_c(m) \triangleq (b_m < P_{r_i}) \cup [(b_m \geq
P_{r_i}) \cap (P g_i/N_0 < T)]$ with probability
\begin{align}\label{eq:P_Ac}
\Pr[A_c(m)] &= F_{h_i}\bp{ \tfrac{T}{b_{m}} }\nonumber\\
& \quad +\BB{1-F_{h_i}\bp{ \tfrac{T}{b_{m}} }}F_{g_i}\bp{ \tfrac{TN_0}{P} }.
\end{align}

\subsubsection{$m=n$ for $0 \leq m < L+1$}%
The relay battery level remains unchanged if the relay $r_i$ is in the charging
mode but the collected energy does not increase the battery level with
probability
\begin{align}\label{eq:Q1}
p_{m,m} &= \Pr[A_c(m) \cap (b_m \leq b_m + P g_i \kappa < b_{m+1} ] \nonumber\\
&= \Pr[ (b_m < P_{r_i}) \cap (g_i < \tfrac{b_1}{P \kappa })] \nonumber\\
& \quad + \Pr[ \bp{ (b_m \geq P_{r_i}) \cap ( \tfrac{P g_i}{N_0} < T) } \cap (g_i < \tfrac{b_1}{P \kappa} )] \nonumber \\
&= F_{g_i} \bp{\tfrac{b_1}{P \kappa} } F_{h_i}\bp{ \tfrac{T}{b_m} } +
\begin{cases}
F_{g_i}\bp{ \tfrac{TN_0}{P} }, & T < \tfrac{b_1}{\kappa},\\
F_{g_i}\bp{ \tfrac{b_1}{P \kappa} }, & T \geq \tfrac{b_1}{\kappa}
\end{cases} \nonumber \\
& \triangleq Q_1(m).
\end{align}

\subsubsection{$m=0, n=L+1$}%
Given the battery is empty, relay $r_i$ must be in the charging mode. The
probability that the relay battery becomes fully charged is identical to
\begin{align}\label{eq:Q2}
p_{0,L+1} &= \Pr[ P g_i \kappa \geq B] = 1-F_{g_i}\bp{ \tfrac{\alpha}{\kappa} } \nonumber\\
& \triangleq Q_2(0,L+1).
\end{align}

\subsubsection{$m=0,~0 < n < L+1$}%
In this case relay $r_i$ with an empty battery harvests energy from the
received signal such that its battery becomes partially charged. This
transition probability is equal to
\begin{align}\label{eq:Q3}
p_{0,n} &= \Pr[ b_n \leq P h_i \kappa < b_{n+1} ] = F_{g_i}\bp{ \tfrac{b_{n+1}}{P\kappa} }
- F_{g_i}\bp{ \tfrac{b_{n}}{P\kappa} } \nonumber\\
& \triangleq Q_3(0,n).
\end{align}

\subsubsection{$0<m\leq L+1,~n<m$}%
The relay battery level is reduced from level $m$ to level $n$ only if the relay is in the forwarding mode. Hence, the transition probability can be found as
\begin{align}\label{eq:Q4}
p_{m,n} &= \Pr[ A_f(m) \cap (b_n \geq b_m-P_{r_i} < b_{n+1}) ] \nonumber \\
&= \BB{1-F_{g_i}\bp{ \tfrac{TN_0}{P} } } \BB{F_{h_i}\bp{ \tfrac{T}{b_{m}-b_{n+1}} } - F_{h_i}\bp{ \tfrac{T}{b_m-b_n} } }\nonumber\\
& \triangleq Q_4(m,n).
\end{align}

\subsubsection{$0<m \leq L+1, n=L+1$}%
The partially charged relay battery becomes fully charged if the relay is in
the charging mode and the amount of harvested energy exceeds the remaining
space of the battery, whose probability is given by
\begin{align}\label{eq:Q5}
p_{m,L+1} &= \Pr[A_c(m) \cap (P g_i \kappa \geq B-b_m)] \nonumber \\
&= F_{h_i}\bp{ \tfrac{T}{b_m} } \BB{ 1- F_{g_i}\bp{ \tfrac{\alpha P - b_m}{P \kappa} } } \nonumber\\
& \quad + \begin{cases}
0, & T \leq \tfrac{ \alpha P - b_m}{ \kappa}, \\
F_{g_i}\bp{ \tfrac{TN_0}{P} } - F_{g_i}\bp{ \tfrac{\alpha P - b_m}{P \kappa} },
&  T > \tfrac{ \alpha P - b_m}{ \kappa}
\end{cases} \nonumber \\
& \triangleq Q_5(m,L+1).
\end{align}

\subsubsection{$0 < m < n < L+1$}%
This corresponds to the case that the non-empty battery remains not full after
harvesting the energy, which takes place with probability
\begin{align}\label{eq:Q6}
&p_{m,n} = \Pr[ A_c(m) \cap (b_n < b_m+P g_i \kappa < b_{n+1}) ] \nonumber \\
&= F_{h_i}\bp{ \tfrac{T}{\kappa} } \BB{ F_{g_i}\bp{ \tfrac{b_{n+1}-b_n}{P \kappa} }
- F_{g_i}\bp{ \tfrac{b_n-b_m}{P \kappa} }} \nonumber \\
& + \begin{cases}
0, & T < \tfrac{ b_n-b_m }{\kappa}, \\
F_{g_i}\bp{ \tfrac{T N_0}{P} } - F_{g_i}\bp{ \tfrac{b_n-b_m}{P \kappa} },
& \tfrac{ b_n-b_m }{\kappa} \leq T < \tfrac{ b_{n+1}-b_m }{\kappa}, \\
F_{g_i}\bp{ \tfrac{b_{n+1}-b_m}{P \kappa} } - F_{g_i}\bp{ \tfrac{b_n - b_m}{P
\kappa} }, & T \geq \tfrac{ b_{n+1}-b_m }{\kappa}
\end{cases} \nonumber \\
& \triangleq Q_6(m,n).
\end{align}

\subsubsection{$m = n = L+1$}%
This case arises only when the relay is in the charging mode with probability
\begin{align}\label{eq:Q7}
p_{L+1,L+1} &= \Pr[ A_c(L+1) ]  \triangleq Q_7(L+1),
\end{align}
where $\Pr[ A_c(L+1) ]$ has been given in (\ref{eq:P_Ac}).

\subsection{Transition Probability Matrix}
Based on the state transition probabilities obtained in the previous subsection, the transition probability matrix of the DTMC can be constructed. Since the number of states grows exponentially with the number of relays, a systematic approach is provided below to facilitate the construction of the transition probability matrix.
\begin{description}
\item[Step 1:]~Identify all the possible combinations of the forwarding set $\mathcal{F}(s)$. Given $N$ relays, there are $2^N$ different configurations of $\mathcal{F}(s)$.
\item[Step 2:]~For each $\mathcal{F}(s)$, find the associated transition probabilities. For example, if $L=1$ and $N=2$, there will be $2^2$ combinations of $\mathcal{F}(s)$. Consider $\mathcal{F}(s)=\{r_1\}$, one of the potential state transitions is from $(1,0)$ to $(0,0)$ where $r_1$'s battery level is reduced by one level with probability $Q_4(1,0)$ and $r_2$'s battery remains empty with probability $Q_1(0)$.
\item[Step 3:]~Once the transition probability matrix $\mathbf{P}$ is obtained, the steady-state probabilities can be obtained by
solving the balanced equation $\boldsymbol{\pi} \mathbf{P} =
\boldsymbol{\pi}$ along with the normalized condition
$\sum_{m=1}^{(L+2)^N} \pi_m=1$
\end{description}

\section{Results and Discussions}\label{sec:results}
This section presents numerical results to demonstrate the performance of the
proposed BARS scheme. Unless specified, the following parameters are used
throughout this section: $R=1$ bits/Sec/Hz, $N_0=1$, and
$\bar{g}_i=\bar{h}_i=1$. In simulations, all relay batteries are set to be full
initially. Each curve in the figure is obtained from $10^6$ independent runs.
Besides, we use ``Theo'' and ``Sim'' to indicate the theoretical and simulation
results, respectively.

\begin{figure}[!t]
\centering
\includegraphics[width=1.0\columnwidth]{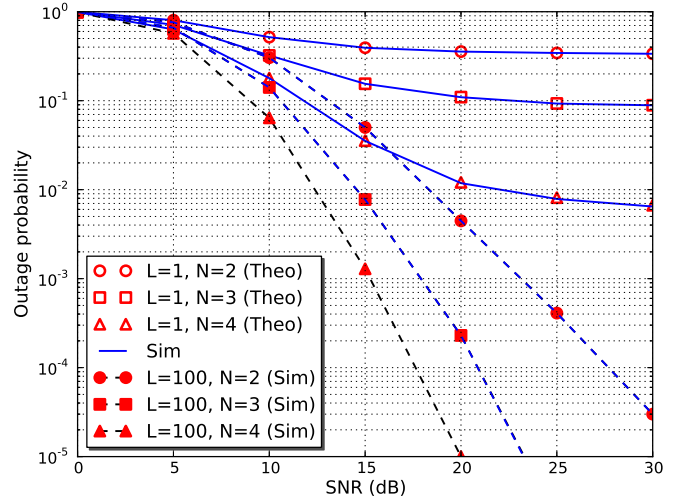}
\caption{Outage probability of BARS under varied $L$ and $N$ with fixed
$\kappa=0.5$.} \label{fig:pout-vs-L}
\end{figure}

Fig.~\ref{fig:pout-vs-L} shows the outage probability of BARS versus SNR under
different number of relays $N$ and the number of battery quantization levels
$L$. Here we fix the battery scaling factor $\alpha=1$ and the energy
conversion efficiency $\kappa=0.5$. For $L=1$, both theoretical and simulation
results are included while only simulation results are shown for $L=100$. It
can be seen that the theoretical results agree with the simulated ones well.
When $L=1$, the outage probability decreases with $N$ but incurs a severe error
floor at high SNR. This is because a small $L$ implies a lossy quantization
interval. In this case, the amount of stored energy can hardly reach the
quantization threshold and thus the battery is often at the low
level. This problem can be resolved by increasing $L$ that in turns reduces the
risk of no relays eligible to help. As $L$ increases to 100, the slope of the
outage probability curve is equal to the number of relays $N$, which suggests
that BARS can fully explore diversity gain provided with sufficient battery
quantization levels.

\begin{figure}[!t]
\centering
\includegraphics[width=1.0\columnwidth]{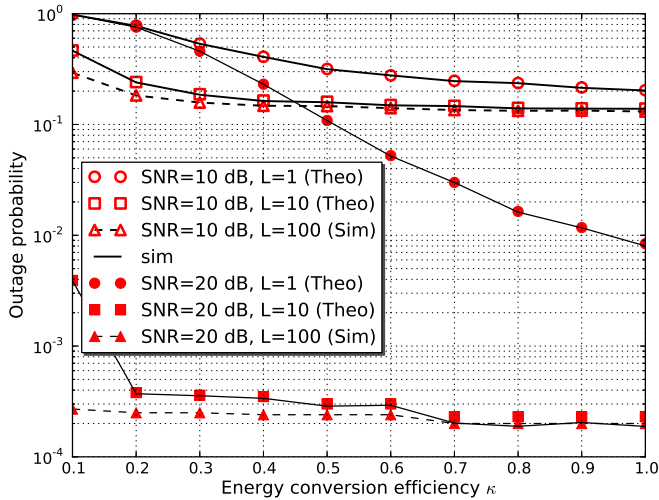}
\caption{Outage probability of BARS vs. energy conversion efficiency$\kappa$
under varied $L$ and SNR for $N=3$.} \label{fig:pout-vs-c}
\end{figure}

Fig.~\ref{fig:pout-vs-c} investigates the impact of energy conversion
efficiency $\kappa$ for $N=3$. Intuitively, the outage probability decreases
with increasing $\kappa$, but the decreasing trend is less significant when $L$
is large. This is because a larger $L$ implies a finer granularity of the
battery level such that the battery status is not sensitive to $\kappa$. For
$L=10$ and 100, the outage performance of BARS is nearly saturated at
$\kappa=0.5$, which is about the practical value of most RF energy harvesters.
In other words, the room for further improvement by more sophisticated energy
harvester might not be as notable as expected.

\begin{figure}[!t]
\centering
\includegraphics[width=1.0\columnwidth]{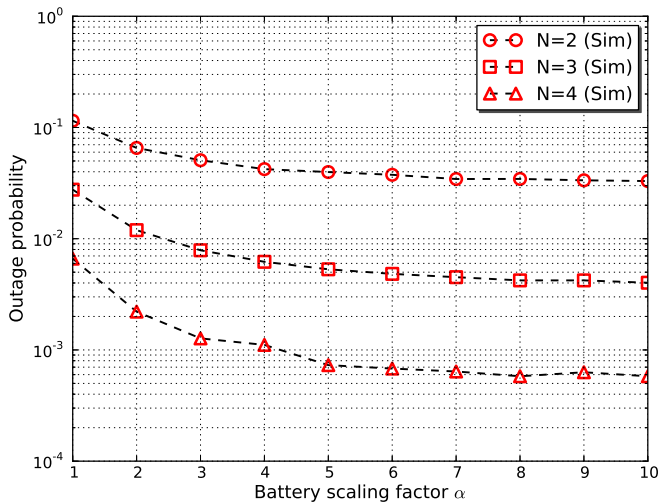}
\caption{Outage probability of BARS vs. battery scaling factor $\alpha$ under
varied $N$ with $L=100$, $\kappa=0.5$, SNR=20 dB, and $R=2$ bits/sec/Hz.}
\label{fig:pout-vs-a}
\end{figure}

The impact of battery size to the outage performance is explored in
Fig.~\ref{fig:pout-vs-a} by varying the battery scaling factor $\alpha$. Here,
we set $L=100$, $\kappa=0.5$, $\text{SNR}=20$ dB, $R=2$ bits/sec/Hz, and only
simulation results are shown. It follows the intuition that increasing the battery size helps to
reduce the outage probability, but the gain becomes diminished when
$\alpha>0.6$, regardless of the number of relays $N$. This can be explained by
the fact that when the battery is larger than a certain degree, it is less
likely to be fully charged given a limited energy conversion efficiency. On the
other hand, the impact of battery size is more significant when $N$ is larger,
as a consequence of selection diversity.

\begin{figure}[!t]
\centering
\includegraphics[width=1.0\columnwidth]{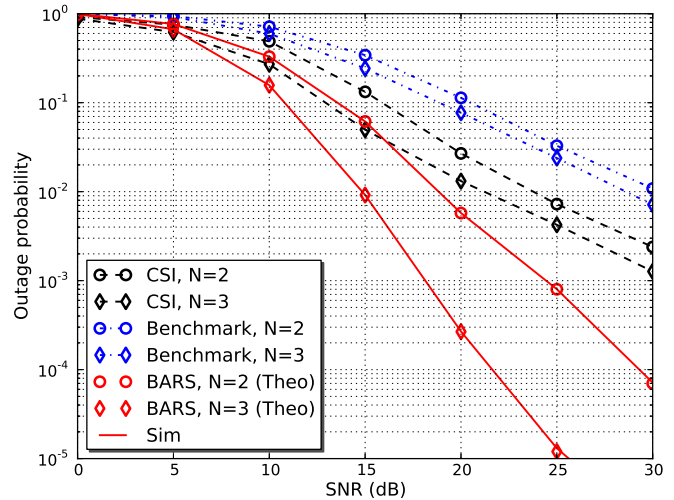}
\caption{Comparison of outage probabilities for BARS and CSI-based relay
selection with $L=10$ and $\kappa=0.5$.} \label{fig:pout-comparison}
\end{figure}

Finally, we compare the performance of BARS with the CSI-based relay selection
scheme in~(\ref{eq:csi}) and the benchmark scheme in (\ref{eq:benchmark}). In
Fig.~\ref{fig:pout-comparison}, we fix $\kappa=0.5$ and consider $N=2$ and 3.
One can see the significant improvement of BARS over the CSI-based and the
benchmark relay selection schemes, both failing to achieve full diversity gain. We note that the diversity order of BARS shown in the graph is slightly
less than $N$ primarily because a small quantization resolution of the battery
is used ($L=10$). BARS can achieve full diversity order provided with $L=100$,
as shown in Fig.~\ref{fig:pout-vs-L}.

\section{Conclusion}\label{sec:conclusion}
In this paper, we proposed a relay selection scheme called BARS for EH relays.
BARS differs from traditional relay selection schemes based on CSI in that it
takes into account the battery status of relays in order to prevent selecting
the relay lacking of energy to forward the source signal, a key factor
that deteriorates the performance of EH relays with finite battery. In BARS,
the relays that can decode the source information and have sufficient power to
transmit are defined as the forwarding set. In this set, the best relay is
chosen as the one that has the least harvested energy, depending on the
source-relay channel condition and the energy conversion efficiency of the
energy harvester. Such a selection allows the network to collect the largest
amount of harvested energy per source transmission. The performance of BARS is
analyzed theoretically based on a discretized battery model. Our results reveal that BARS
achieves full diversity order and significantly outperforms the traditional
CSI-based relay selection scheme, which fails to fully exploit diversity gain
provided by multiple EH relays.


%
%
%

\bibliographystyle{IEEEtran}
\bibliography{madmf}

\end{document}